\documentclass{aa}

\usepackage{graphicx}
\usepackage{txfonts}

\usepackage[american]{babel}
\usepackage[T1]{fontenc}
\usepackage[applemac]{inputenc}
\usepackage{palatino}
\usepackage{times}
\usepackage{epsfig}
\usepackage{amssymb}
\usepackage{lscape}
\usepackage{color}

\newcommand{\comments}[1]{}

\begin{document}

\title{Simulation of optical interstellar scintillation}
\author{F.~ Habibi\inst{1}$^,$\inst{2},
M.~Moniez\inst{1},
R.~Ansari\inst{1},
S.~Rahvar\inst{3}$^,$\inst{4}
}
\institute{
Laboratoire de l'Acc\'{e}l\'{e}rateur Lin\'{e}aire,
{\sc IN2P3-CNRS}, Universit\'e de Paris-Sud, B.P. 34, 91898 Orsay Cedex, France
\and
School of Astronomy, Institute for Research in Fundamental Sciences (IPM), PO Box 19395-5531,Tehran, Iran
\and
Department of Physics, Sharif University of Technology
PO Box 11365-9161, Tehran, Iran
\and
Perimeter Institute for Theoretical Physics, 31 Caroline Street North, Waterloo, Ontario N2L 2y5, Canada
}

\date{\today}

%\begin{document}
%\maketitle
\abstract{}
%\begin{abstract}
{
Stars twinkle because their light propagates through the atmosphere.
The same phenomenon is expected on a longer time scale
when the light of remote stars
crosses an interstellar turbulent molecular cloud, but it has never been
observed at optical wavelengths.
The aim of the study described in this paper is to fully simulate the scintillation
process, starting from the molecular cloud description as a fractal object,
ending with the simulations of fluctuating stellar light curves.
}
{
Fast Fourier transforms are first used to simulate
fractal clouds.
%Their effect on the propagation of background stars'
%light have been calculated through the Fresnel integral.
Then, the illumination pattern resulting from the crossing of background
star light through these refractive clouds is calculated
from a Fresnel integral that also uses fast Fourier transform techniques.
Regularisation procedure and computing limitations are discussed,
along with the effect of spatial and temporal coherency (source size and
wavelength passband).
}
{
We quantify the expected modulation index of stellar
light curves as a function of the turbulence strength
--characterised by the diffraction radius $R_{diff}$-- and
the projected source size, introduce the timing aspects, and
establish connections between the light curve
observables and the refractive cloud.
We extend our discussion to clouds with different structure functions from
Kolmogorov-type turbulence.
%Finaly, we propose an observational strategy to discover the first optical
%scintillation effects.
}
{
Our study confirms that current telescopes of $\sim 4\, m$ with
fast-readout, wide-field detectors have the capability of discovering the
first interstellar optical scintillation effects.
We also show that this effect should be unambiguously distinguished
from any other type of variability
%The observing strategy has been refined through the use of the
%simulation, and we show that
through the observation of desynchronised
light curves, simultaneously measured by two distant telescopes.
}
%\end{abstract}

\titlerunning{Simulation of Optical Interstellar Scintillation}
\authorrunning{Habibi, Moniez, Ansari, Rahvar}
\maketitle

%\section{Introduction}
%\begin{figure*}[!ht]
%\begin{minipage}[b]{0.48\linewidth}
%\centering
%\includegraphics[width=8.cm]{phi100-3.eps} 
%%\includegraphics[width=8.cm]{/Users/habibi/Documents/These/Fig/phi100-3.eps} 
%\caption[] 
%{\it
%Simulation of the phase screen fluctuations with  $N_{x}\times N_{y} =
%10,000times 10,000$, $\Delta_1$ = 32.6 km 
%and $R_{diff} $ =100 km. Phase fluctuations are given in radian (grey scale).
%}
%\label{cloud}
%\end{minipage}
%\hspace{.4cm}
%\begin{minipage}[b]{0.48\linewidth}
%\centering{\it
%The source is located in the ($x_2,y_2$) plane, the screen contains the diffusive structure, and the observer is 
%located in the ($x_0,y_0$) plane. $A_1(x_1,y_1)$ and $A'_1(x_1, y_1)$ are the amplitudes before and after screen
%crossing.
%}
%\label{geom}
%\end{figure}
\section{Introduction}

This paper is a companion paper to the observational results published
in \cite{habibietal} (2011b), and it focusses on the simulation of the
scintillation effects that were searched for.
Cold transparent molecular clouds are one of the last possible
candidates for the missing baryons of cosmic
structures on different scales (\cite{fractal}, \cite{pfre}
and \cite{mcetal}). Since these hypothesised clouds do not
emit or absorb light, they are invisible
for the terrestrial observer, so we have to investigate indirect
detection techniques. Our proposal for detecting such transparent clouds
is to search for the scintillation %effect on the light curves
of the stars located behind the transparent medium, caused
by the turbulence of the cloud (\cite{Moniez} and \cite{habibietal} 2011b).
%The scintillation of compact radio-sources is well established in
%radioastronomy (\cite{lyne} and \cite{narayan92})
The objective of this technical paper is to describe the way we can
connect observations to scintillation parameters through a
realistic simulation. We used these connections in the companion
paper (\cite{habibietal} 2011b) to establish constraints both from null
results (towards SMC) and from observations pointing to a possible
scintillation effect (towards nebula B68).
Similar studies of propagation through a stochastic medium followed
by Fresnel diffraction have been made by \cite{colesetal}
and for use in radio-astronomy by \cite{hamidouche}.

We first introduce the notations and the formalism in section \ref{sec:formalism}.
Then we describe the different stages of the simulation pipeline
up to the production of simulated light curves in section \ref{sec:simul}.
We study the observables that can be extracted from the light curve
of a scintillating star, and in particular, we
check the expected modulation amplitude properties
in section \ref{sec:observables}.
The discussion is extended to non-Kolmogorv turbulence cases in
section \ref{sec:turbulence}.
%We describe the use the simulation pipeline in order to optimise the
%observational strategy for discovering
%scintillating stars in section \ref{sec:discussion}, and indicate
%some perspectives in the conclusion.
In section \ref{sec:discussion} we use the results from 
the simulation pipeline to optimise the
observational strategy for discovering
scintillating stars, and indicate
some perspectives in the conclusion.

Complementary information on observations made with the ESO-NTT
telescope and on the analysis based on the present simulations are to
be found in our companion paper (\cite{habibietal} 2011b).

\section{Basic definitions and formalism}
\label{sec:formalism}
\begin{figure}[!ht]
\begin{center}
\includegraphics[width=9.cm]{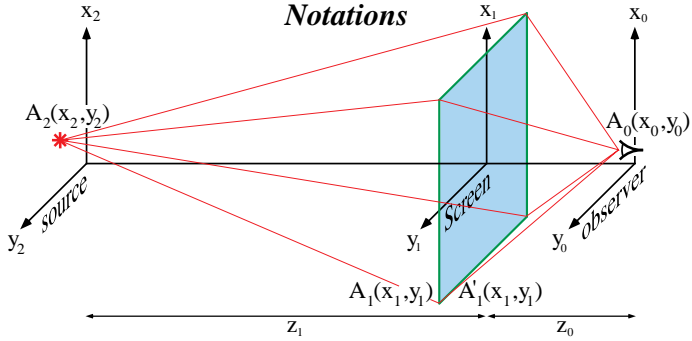}
\end{center}
\caption[] 
{\it
Geometric configuration. The source is located in the ($x_2,y_2$) plane, the screen contains the refractive structure, and the observer is 
located in the ($x_0,y_0$) plane. $A_1(x_1,y_1)$ and $A'_1(x_1, y_1)$ are the amplitudes before and after screen
crossing.
}
\label{geom}
\end{figure}
The formalism described in this section has been inspired and
adapted from the radioastronomy studies (\cite{narayan92} 1992).
But at optical wavelength, the scintillation is primarily due to the refraction
through dense clouds of $H_2+He$ instead of the interaction with the
ionised interstellar medium.
The origin of the stochastic phase fluctuation experienced
by the electromagnetic wave when crossing the refractive medium
is the phase excess induced by the stochastic fluctuation of
the column density due to the turbulence (\cite{Moniez}):
\begin{eqnarray}
\label{phiNl}
\phi(x_1,y_1) &=& \frac{(2 \pi)^2 \, \alpha}{\lambda} Nl(x_1,y_1)
\end{eqnarray}
where $x_1$ and $y_1$ are the coordinates in the cloud's plane, perpendicular to the sightline 
(see figure \ref{geom}). Here $\phi(x_1,y_1)$ is the 
phase delay induced to the wavefront after crossing the cloud,
$Nl(x_1,y_1)$ is the cloud column density of $H_2$ molecules plus He
atoms along the line of
sight, $\alpha$ is the medium polarisability, and $\lambda$ the wavelength.
The phase delay here scales with $\lambda^{-1}$, in contrast
to the radioastronomy where it scales with $\lambda$.
Since other sources of
phase delay, such as the geometrical delay induced by scattering from
cloud inhomogeneities, are negligible, the thin screen
approximation can be used, and the cloud can be considered as a 2D
scattering screen whose optical properties are 
mapped by the phase screen $\phi(x_1,y_1)$.
\begin{figure*}[!ht]
\begin{minipage}[b]{0.48\linewidth}
\centering
\includegraphics[width=8.cm]{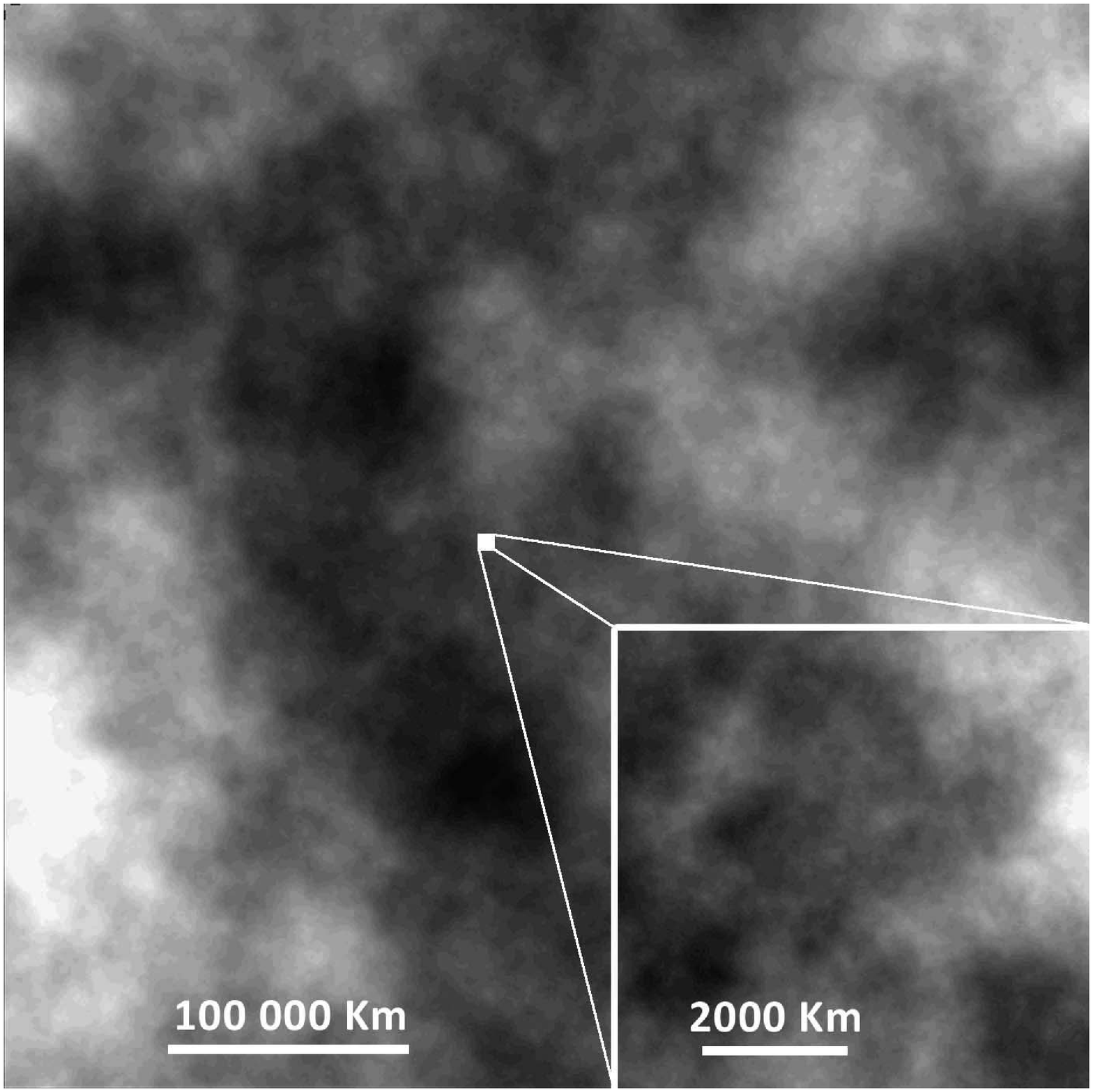} 
\caption[] 
{\it
%Simulation of the phase screen fluctuations with  $N_{x}\times N_{y} =
%10,000\times 10,000$, $\Delta_1$ = 32.6 km 
%and $R_{diff} $ =100 km. Phase variations are given in radian (grey scale).
The phase-delay variations near the average
for a simulated refractive screen with  $N_{x}\times N_{y} =
20,000\times 20,000$ pixels, $\Delta_1$ = 22.6 km, 
and $R_{diff} $ =100 km.
The grey scale ranges between $\pm 50\times 2\pi\, rad$ (clear
regions correspond to an excess of phase with respect to the average).
The zoom (inset) illustrates the self-similarity of the simulated screen
(grey scale amplitude of $5\times 2\pi\, rad$).
}
\label{cloud}
\end{minipage}
\hspace{.4cm}
\begin{minipage}[b]{0.48\linewidth}
%\centering {\it
%The source is located in the ($x_2,y_2$) plane, the screen contains the diffusive structure, and the observer is 
%located in the ($x_0,y_0$) plane. $A_1(x_1,y_1)$ and $A'_1(x_1, y_1)$ are the amplitudes before and after screen
%crossing.
%}
%\label{geom}
%\end{figure}
\includegraphics[width=8.cm]{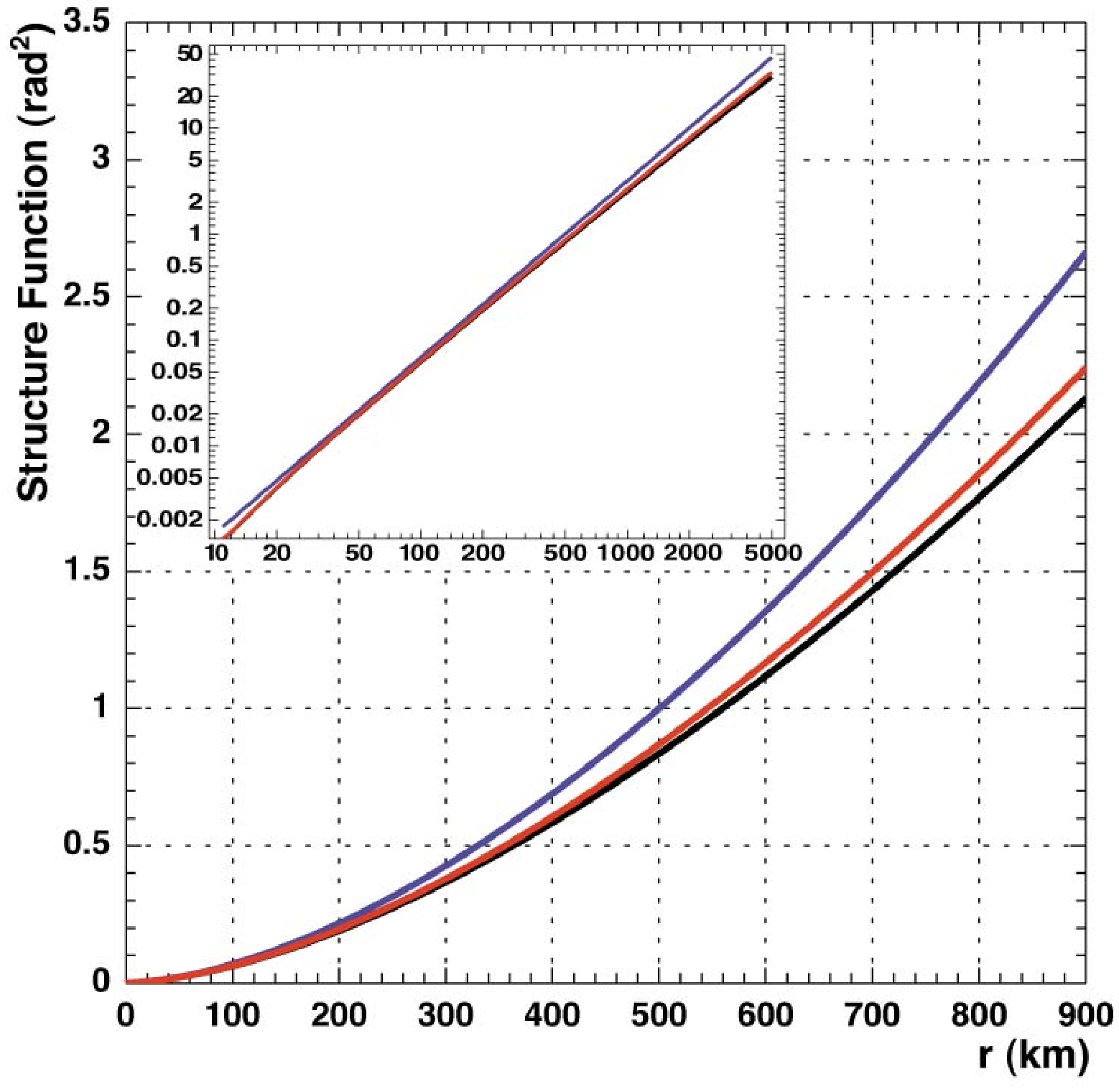}
\caption[] 
{\it
  Phase structure functions $D_\phi(r)$ for a phase screen with $R_{diff}$ = 500 km.
  Blue line is the initial (theoretical) structure function. 
  Red line is reconstructed from one of the realisations of the phase screen through simulation. 
  The black curve is obtained from the numerical integration
  of the initial phase spectral density sampled as in the simulation. 
}
\label{sfrec}
\end{minipage}
\end{figure*}
The statistical properties
of the phase screen are described by the phase structure function
$D_\phi(x_1,y_1)$. By assuming an isotropic turbulence
(\cite{narayan92} 1992):
\begin{eqnarray}
&&D_\phi(x_1,y_1)= D_\phi(r) \nonumber \\
&&=\langle\left[\phi(x_1\!+\!x_1^{\,\prime},y_1\!+\!y_1^{\,\prime})\!-
\phi(x_1^{\,\prime},y_1^{\,\prime})\right]^2\rangle_{(x_1^{\,\prime},y_1^{\,\prime})}\!=\! \left[\frac{r}{R_{diff}}\right]^{\beta-2}\!\!\!, 
\label{dphi}
\end{eqnarray}
where the first expression is averaged over the plane positions
$(x_1^{\,\prime},y_1^{\,\prime})$,
$r=\sqrt{x_1^2+y_1^2}$, $\beta$ is the
turbulence exponent --equals 11/3 for Kolmogorov turbulence-- and the
diffraction radius, $R_{diff}$, is the transverse distance on the phase
screen where the phase changes by one radian on average. The diffraction radius can be
expressed in terms of the cloud parameters (\cite{habibietal} 2011b);
assuming Kolmogorov turbulence it is given by
\begin{eqnarray}
R_{diff}(\lambda)\! =\! 263{km}\left[\frac{\lambda}{1 \mu {m}}\right]^{\frac{6}{5}} \left[\frac{L_z}{10 {AU}}\right]
	^{-\frac{3}{5}} 
	\left[\frac{L_{out}}{10 { AU}}\right]^{\frac{2}{5}}
        \left[\frac{\sigma_{3n}}{10^9
            {cm}^{-3}}\right]^{-\frac{6}{5}}\!\!\!\! ,
%R_{diff}\! =\! 744{km}\left[\frac{\alpha}{\alpha(H_2/He)}\right]^{-\frac{6}{5}}\left[\frac{\lambda}{1 \mu {m}}\right]^{%\frac{6}{5}} \left[\frac{L_z}{10 {AU}}\right]
%	^{-\frac{3}{5}} 
%	\left[\frac{L_{out}}{10 { AU}}\right]^{\frac{2}{5}} \left[\frac{\sigma_{3n}}{10^9 {cm}^{-3}}\right]^{-\frac{6}{5}},
\label{relrdiffout}
\end{eqnarray} 
where $L_z$ is the cloud size along the sightline, $L_{out}$ is the turbulence outer scale, and 
$\sigma_{3n}$ is the dispersion of the volumic number density in the
medium\footnote{
A cloud column of width $L_z$ can include several turbulent structures
with outer scale $L_{out}$. The direct relation of $L_{out}$ with the
turbulence strengh explains why $R_{diff}$ increases
in conjunction with this parameter.
By contrast, since the column density increases with $L_z$,
then the refraction also increases, thus decreasing $R_{diff}$.}.
Here we assume the gas to be a mixing of  $H_2/He$ with $24\%$
$He$ by mass (corresponding to the primordial abundances) and
therefore $<\alpha>=0.720\times 10^{-30} m^3$.
In this expression, the cloud parameters 
are scaled to the values given by the Pfenniger-Combes model for the
clumpuscules (the tiniest cloudlets of the 
molecular cloud). In the NIR band, the diffraction
radius of a typical clumpuscule is expected to be $R_{diff} \sim $ 500
km.

The phase statistics of the screen can be equivalently described in Fourier space by
the phase spectral density: % $S_\phi(q_x,q_y)$:
\begin{eqnarray}
S_\phi(q_x,q_y) = S_\phi(q) = \frac{R_{diff}^2}{2\,(2\pi)^{\beta-1}\,f(\beta)}\,(R_{diff}\,q)^{-\beta},
\label{spectre}
\end{eqnarray}
where Fourier coordinates $q_x$ and $q_y$ have inverse length dimension, $q$ = $\sqrt{q_x^2+q_y^2}$,  
and $f(\beta) = \frac{2^{-\beta}\, \beta \,\Gamma(-\beta/2)}{\Gamma(\beta/2)}$
%$f(\beta)$
is a constant.

After crossing the cloud,
the distorted wavefront of a {\it point source} propagates toward the observer and produces
an illumination pattern on the observer's plane given by   
\begin{eqnarray}
\label{i0}
I_0(x_0,y_0) &=& \frac{L_s}{z_1^2}\,h(x_0,y_0),
\end{eqnarray}
%where $x_0$ and $y_0$ are the coordinates in the observer plane,
where $I_0(x_0,y_0)$ is the light
intensity on the observer's plane, $L_s$ is the source luminosity,
$z_1$ is the source-screen distance (see Fig. \ref{geom}), and 
$h(x_0,y_0)$ is given by the Fresnel-Huygens diffraction integral
after considering the Fresnel and the stationary phase approximations
(\cite{Born}, \cite{Moniez}):
\begin{eqnarray}
\label{hxy}
%h(x_0,y_0) &=& |\frac{1}{2\pi R_F^2} \,\iint_{-\infty}^{+\infty} e^{i\phi(x_1,y_1)} 
%               e^{i\frac{(x_0-x_1)^2+(y_0-y_1)^2}{2 R_F^2}}dx_1 dy_1|^2,            
h(x_0,y_0) &=& \Big\vert\frac{1}{2\pi R_F^2} \,\iint_{-\infty}^{+\infty} e^{i\phi(x_1,y_1)} 
               e^{i\frac{(x_0-x_1)^2+(y_0-y_1)^2}{2 R_F^2}}dx_1
               dy_1\Big\vert ^2,          
\end{eqnarray}
where $R_F$ = $\sqrt{\lambda \, z_0 /2\pi}$ is the Fresnel
radius\footnote{This definition assumes that $z_0<<z_1$. In the general
case, $z_0^{-1}$ should be replaced by $z_0^{-1}+z_1^{-1}$.}
and $z_0$ is the screen-observer distance. The Fresnel radius
can be expressed as
\begin{eqnarray}
R_F &=& 2214 \, km \left[\frac{\lambda}{1 \mu m}\right]^{\frac{1}{2}}
\, \left[\frac{z_0}{1 kpc}\right]^{\frac{1}{2}}.
\label{Rfresnel}
\end{eqnarray}
%The Fresnel radius
%for a clumpuscule located
At the typical distance of a halo object ($\sim$ 10 kpc),
$R_F \sim$ 7000 km for NIR wavelengths.
Because of the motion of the cloud with respect to the Earth-source
line-of-sight, the illumination pattern sweeps the observer plane,
so that a terrestrial observer receives fluctuating intensity light
from the point source. 
This effect, the scintillation, has two different scattering
regimes (\cite{Uscinski}, \cite{Tatarskii}),
a weak regime ($R_{diff} > R_F$) and a strong regime
($R_{diff}<R_F$). 
In the present studies, we concentrate on the strong regime,
which clearly is easier to detect, but some realistic configurations may
involve the intermediate regime studied by \cite{goodman} (2006).
For the strong regime, there are two different modes of flux
variations (\cite{narayan92} 1992, see also \cite{Rickett}, \cite{Rumsey}, \cite{Sieber}). 
The first one is the diffractive mode with length scale
corresponding to the screen's scale of phase variations $R_{diff}$
given by equation (\ref{relrdiffout}).
The resulting "speckles", with typical size of the order of $R_{diff}$,
are shown in figure \ref{pattern}.
The corresponding time scale of the light fluctuations is $t_{diff}$ = $R_{diff} / V_T$:
 \begin{eqnarray}
&&t_{diff}(\lambda) =   \\ 
&&\!\!\!\!\! 2.6{s}  \left[\frac{\lambda}{1 \mu {m}}\right]^{\frac{6}{5}} \left[\frac{L_z}{10 {AU}}
	\right]^{-\frac{3}{5}} 
	\left[\frac{L_{out}}{10 {AU}}\right]^{\frac{2}{5}} \left[\frac{\sigma_{3n}}{10^9 {cm}^{-3}}\right]^{-\frac{6}{5}}
	\left[\frac{V_T}{100\, {km/s}}\right]^{-1}, \nonumber
\label{tdiff}
\end{eqnarray} 
where $V_T$ is the sightline relative transverse motion.
Therefore fast flux variations are expected with a typical time scale of
$t_{diff} \sim few\, s$. The second variation mode is the
refractive mode associated to the longer length scale called refraction radius:
\begin{eqnarray}
R_{ref}(\lambda)=\frac{\lambda z_0}{R_{diff}(\lambda)} \sim
30,860\, {km}\left[\frac{\lambda}{1\mu {m}}\right]\left[\frac{z_0}{1\,
    {kpc}}\right]\left[\frac{R_{diff}(\lambda)}{1000\,
    {km}}\right]^{-1}\!\!\!\!\! .
%&&\!\!\!\!\!\!=\!41,500\, {km}\left[\frac{\lambda}{1 \mu {m}}\right]^{-\frac{1}{5}}
%\left[\frac{z_0}{1kpc}\right] \left[\frac{L_z}{10 {AU}}\right]^{\frac{3}{5}} 
%\left[\frac{L_{out}}{10 { AU}}\right]^{-\frac{2}{5}}
%\left[\frac{\sigma_{3n}}{10^9 {cm}^{-3}}\right]^{\frac{6}{5}}.
\label{Rrefraction}
\end{eqnarray}
This natural length scale corresponds to the size, in the observer's
plane, of the diffraction spot from a patch of $R_{diff}(\lambda)$ in the screen's plane.
This is also the size of the region in the screen where most of the
scattered light seen at a given observer's position originates.
Our convention for $R_{ref}$ differs from \cite{narayan92} (1992) by a factor $2\pi$ since
it emerges naturally from the Fourier transform we use for
calculating the illumination pattern (see formula \ref{intens}),
and it also matches
%this size is also close to the typical size of the bright/dark zones
%in the simulated illumination pattern.
the long distance-scale flux variations visible in figure \ref{pattern}.
The corresponding time scale is given by $t_{ref}=R_{ref}/V_T$:
\begin{eqnarray}
t_{ref}(\lambda) \simeq  5.2\, {min.} \left[\frac{\lambda}{1\mu {m}}\right]\left[\frac{z_0}{1\, {kpc}}\right]	\left[\frac{R_{diff}(\lambda)}{1000\, {km}}\right]^{-1}\left[\frac{V_T}{100\, {km/s}}\right]^{-1}\!\!\!\!\! .
\label{em}
\end{eqnarray}

\section{Simulation description}
\label{sec:simul}
In this section, we describe the simulation pipeline with
some numerical tricks, from the
generation of the phase screen induced by turbulent gas, up to the
light versus time curves expected from realistic stars seen through
this gas. The steps in this pipeline are listed below:
\begin{itemize}
\item
The simulation of the refractive medium:
in subsection \ref{sec:phasescreen} we describe the generation of
a phase screen and examine the impact of the
limitations caused by the
sampling and by the finite size of the screen by comparing the
initial (theoretical) and reconstructed diffraction radii.
\item
The computation of the illumination pattern:
in subsection \ref{sec:illpattern}, we first describe
the calculation of the illumination pattern produced on Earth by a point
monochromatic source as seen through the refractive medium.
We explain the technique for avoiding numerical (diffraction) artefacts
caused by the borders of the simulated screen, and discuss the criterion
on the maximum pixel size to avoid aliasing effects.
The pattern computation is then generalised to extended polychromatic sources.
\item
The light curve simulation:
we describe in subsection \ref{sec:simlightcurve} the simulation
of the light fluctuations with time at a given position
induced by the motion of the refractive medium with respect
to the line of sight.
\end{itemize}
\subsection{Simulation of the phase screen}
\label{sec:phasescreen}
\begin{figure*}[!ht]
\begin{minipage}[b]{0.48\linewidth}
\centering
\includegraphics[width=8.cm]{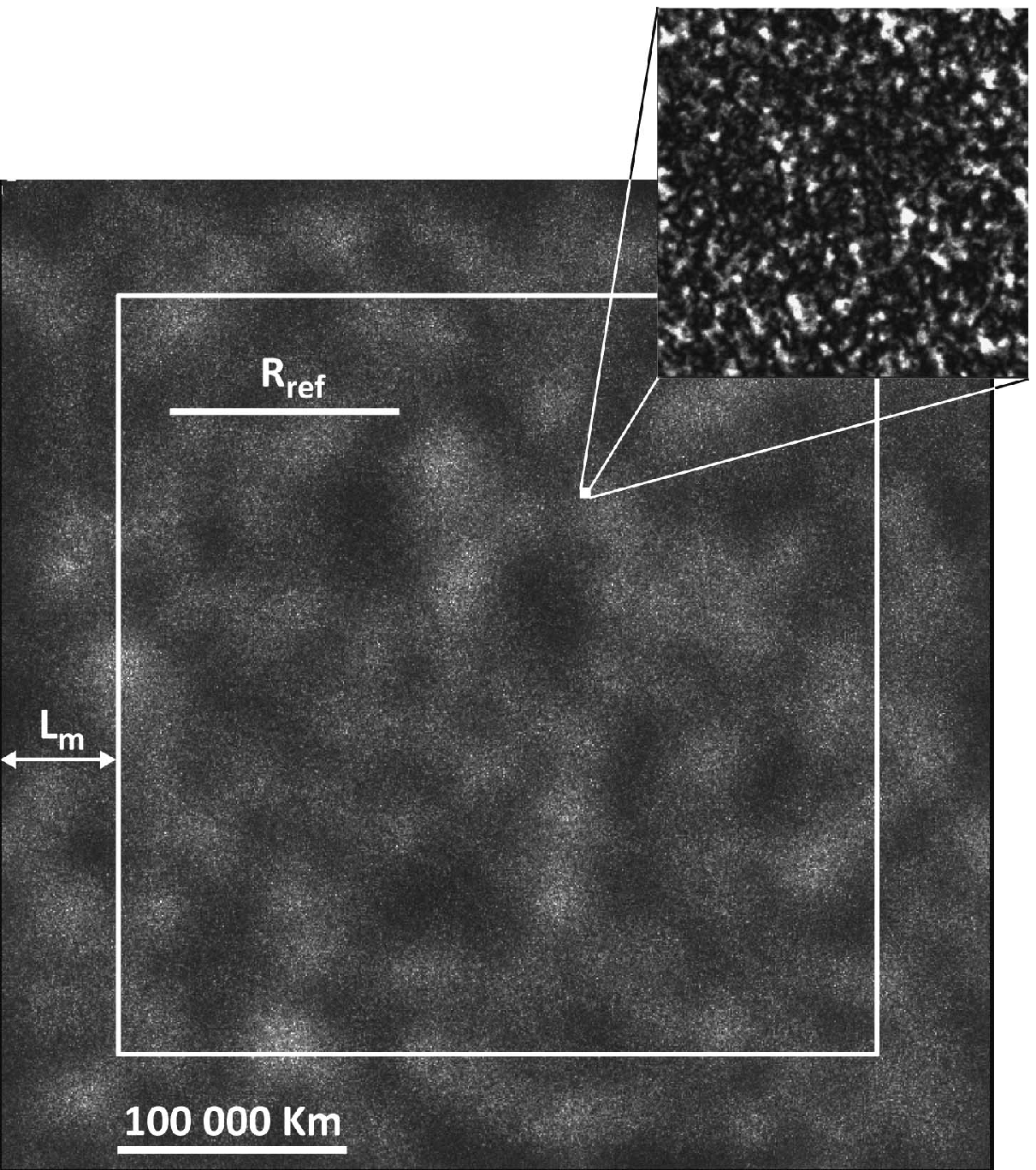}

\caption[] 
{\it
        Typical illumination pattern from a point-like source. Here
        $R_{diff}$ =100 km, the screen is at $z_0=160\, pc$,
        $\lambda=2.16\mu m$, then
$R_{F}$= 1300 km and $R_{ref}$ = 106000 km. 
	  The typical length scale of the small-scale speckles is $R_{diff}$, and the scale of the larger 
	  structures is $R_{ref}$. The white square shows our
          fiducial zone with a margin of
	  $L_m=R_{ref}/2$ from the borders. Grey scale range from 0 to
          4 times the mean intensity. The image has $20,000\times
          20,000$ pixels, each with a $22.6\, km$ side.
}
\label{pattern}
\end{minipage}
\hspace{0.4cm}
\begin{minipage}[b]{0.48\linewidth}
\centering
\includegraphics[width=8.cm]{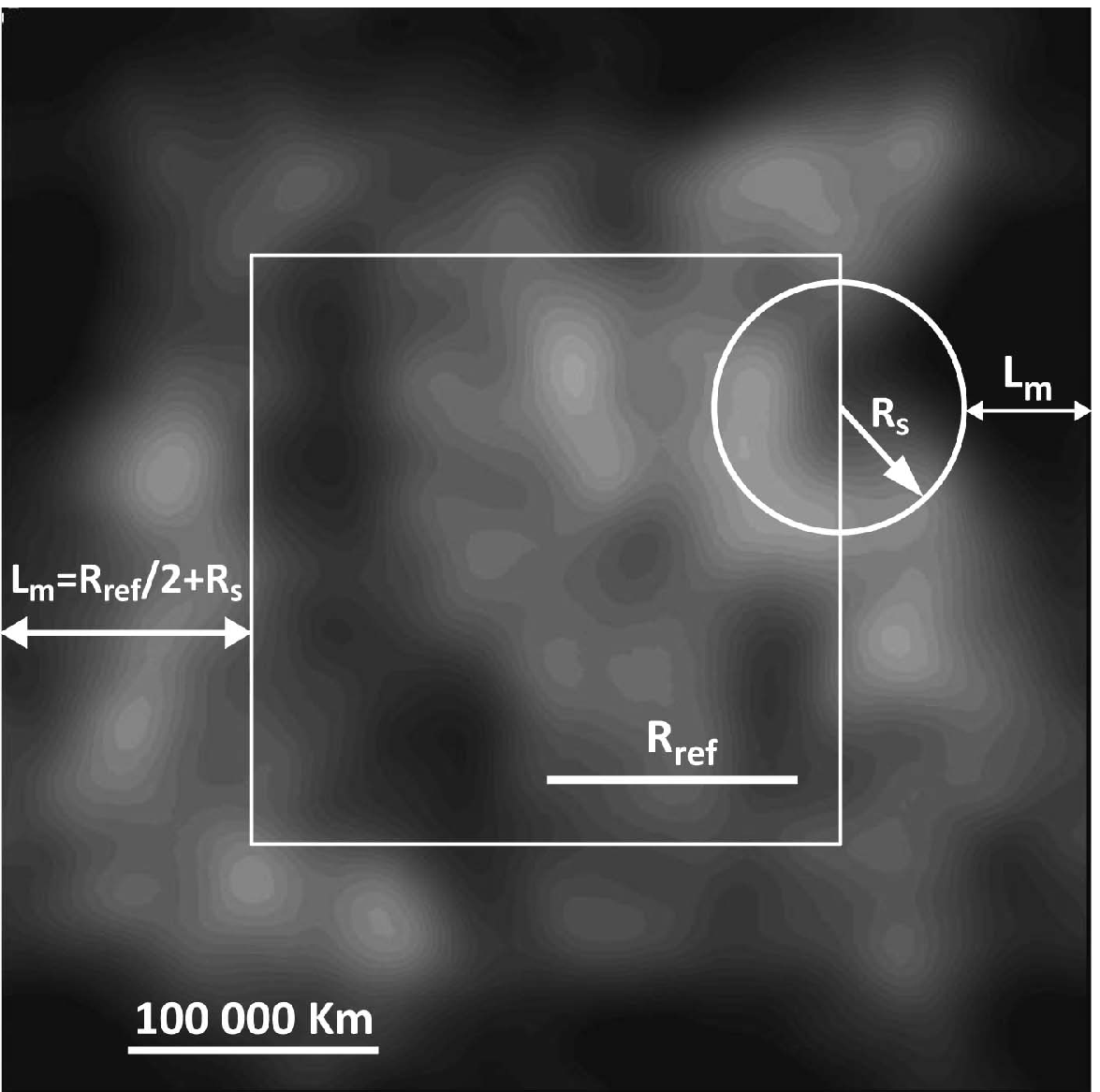}

\caption[] 
{\it
Typical illumination pattern from an extended source
produced through the same screen as in Fig. \ref{pattern}, with
source radius $r_s = 0.5 R_{\odot}$, located at $z_0+z_1=1kpc+160pc$
($R_s \simeq 53,000\, km$). The small-scale speckles are smeared, 
	and only the larger scale fluctuations survive. The white square shows the restricted fiducial zone with
	margin of $L_m=R_{ref}/2+R_s$ from the borders. Grey scale
        ranges $\pm 20\%$ around the mean intensity.
}
\label{extended}
\end{minipage}
\end{figure*}
Numerical realisations of the 2D phase screen ---made of $N_{x}\times
N_{y}$ squared pixels of size $\Delta_1$---
are randomly generated from the phase spectral density $S_\phi(q_x,q_y)$,
determined by the choice of $R_{diff}$ in relation (\ref{spectre}).
Such a phase screen ---with the desired statistical properties---
is obtained from the random realisation of a Fourier transform $F_{\phi}(q_x,q_y)$
in a way that makes the ensemble of such realisations satisfy the relation
\begin{equation}
< | F_{\phi}(q_x , q_y) | ^ 2>_{realisations}=L_x L_y S_\phi(q=\sqrt{q_x^2+q_y^2}),
\label{densite}
\end{equation}
where $L_x$ and $L_y$ are the screen physical size, and the average is
the ensemble averaging over the different realisations.
For each $(q_x^i,q_y^j)$ vector associated to a pixel $(i,j)$, we generate a random
complex number $F_{\phi}(q_x^i,q_y^j)=(f_{Re}^{i,j}+{\bf i}.f_{Im}^{i,j})/\sqrt{2}$ where each
component $f_{Re}^{i,j}$ and $f_{Im}^{i,j}$
spans the Gaussian distribution\footnote{The turbulence is considered
as a Gaussian field} of zero mean, and $L_x L_y S_\phi(q=\sqrt{q_x^{i\,2}+q_y^{j\,2}})$ width. In
this way, relation \ref{densite} is automatically satisfied when
averaging on a large number of such realisations,
since the average of $| F_{\phi}(q_x , q_y) | ^ 2$ for the ensemble of
$(q_x,q_y)$ vectors with the same module $q$
equals $L_x L_y S_\phi(q)$ by construction.
The phase screen $\phi(x_1,y_1)$ is finally obtained by numerically
computing the inverse Fourier transform of this 
phase spectrum $F_\phi(q_x,q_y)$.
%\begin{eqnarray}
%\phi(x_1,y_1) &=& \iint{F_\phi(q_x,q_y)\,e^{2\pi i \, (x_1 q_x + y_1 q_y)}\,dq_x\,dq_y},
%\label{fphi}
%\end{eqnarray} 
Figure \ref{cloud} shows a phase screen
generated %from relation (\ref{fphi})
by assuming Kolmogorov turbulence ($\beta$ = 11/3). 
\subsubsection{Preliminary checks, limitations}
%To check the quality of the simulated phase screen, we
%reconstructed its effective phase structure function and
To check the accuracy of  the numerically generated phase screen
(Fig. \ref{cloud}), we recomputed the
%we directly measured
%its reconstructed $R_{diff}$ (named $R_{diff}^{rec}$), from the
%reconstructed
phase structure function $D_\phi^{\,rec}(r)$ (and consequently
$R_{diff}^{\,rec}$)
%as follows:
%%$D_\phi^{\,rec}(r)$ is computed back from the reconstructed
%%In this purpose,
%first, we compute back the simulated screens' effective
%phase spectrum $S_\phi^{\,rec}(q)$,
from the generated phase Fourier transform
$F_\phi(q_x,q_y)$, and we compared it with the theoretical phase structure function
(Eq. \ref{dphi}).
First, the spectral density is recomputed from the generated
$F_{\phi}(q_x , q_y)$ using relation
\begin{eqnarray}
S_\phi^{\,rec}(q) &=& \frac{\langle|F_\phi(q_x,q_y)|^2\rangle_{q=\sqrt{q_x^2+q_y^2}}}{L_x\,L_y},
\end{eqnarray}
where the average is performed on  the $(q_x,q_y)$
coordinates\footnote{Here again, we infer 
the ergodicity property that allows us to replace the ensemble
averaging with an average on directions from only one realisation.}
spanning the circle of radius $q=\sqrt{q_x^2+q_y^2}$.
The corresponding phase auto-correlation function is then given by Fourier transform:
\begin{eqnarray}
\xi^{\, rec}( \bold{r}) &=& \iint S_\phi^{\,rec}(\bold{q}) \, e^{2\pi i \, \bold{q}.\bold{r}} d\bold{q},  \nonumber \\
\xi^{\, rec}(r) &=& \int_{q_{min}}^{q_{max}} \int_{0}^{2\pi} q \, S_\phi^{\,rec}(q) e^{2\pi i \, q r cos \theta} d\theta \, dq \nonumber \\
		     &=& \int_{q_{min}}^{q_{max}} 2\pi q \, S_\phi^{\,rec}(q) \, J_0(2\pi q r) \, dq,
\label{crec}
\end{eqnarray}
where $J_0$ is the Bessel function.
The recomputed structure function is then given by 
$D_\phi^{\,rec}(r)$ = 2($\xi^{\,rec}(0) - \xi^{\,rec}(r)$), and the
value of $R_{diff}^{\,rec}$ is deduced from its definition
$D_\phi^{\,rec}(R_{diff}^{\,rec}) = 1$.

In figure \ref{sfrec}, we show the 
theoretical phase structure function of a turbulent medium with
$R_{diff}$ = 500 km --for which 
$D_\phi(r=500\,km)$ = 1 by definition--, and the
recomputed (effective) structure function from one of the
realisations of the screen.
From this recomputed function, we find $R_{diff}^{\,rec} \approx 540\, km$ since
$D_\phi^{\,rec}(r \approx 540\,km)$ = 1. To find the origin of the
difference with the input value $R_{diff}=500\,km$, we replaced
$S_\phi^{\,rec}(q)$ in equation (\ref{crec}) by the theoretical spectrum $S_\phi(q)$ sampled as
in the simulation (number of pixels $N_{x}\times N_{y} \sim
14,000\times 14,000$ with pixel size $\Delta_1$ = 28.85 km).
Then we computed the integral (\ref{crec}) numerically with
the same integration limits ($q_{min}, q_{max}$)
\footnote{In 1D: $q_{min} = \frac{1}{N \Delta_1}$ and $q_{max} = \frac{1}{2 \Delta_1}$.} as the
simulation. The integration result differs
only by a few percent from the function recomputed
from the simulated screen. We showed that the black curve approaches the blue curve 
when $q_{min} \rightarrow 0$ and $q_{max} \rightarrow \infty$. This
means that the sampling
%affects the integration domain.
is mainly responsible for the difference between $D_\phi$ and $D_\phi^{\,rec}$.
%As a conclusion,
Since our simulation is limited by the number of pixels, we lose the contributions of the
large and small scales in the recomputed $R_{diff}$.
The only way to push back this limitation 
is to generate larger screens (larger $N_{x}$ and $N_{y}$) with higher resolutions (smaller $\Delta_1$) to cover 
wider interval of spatial frequencies which in return needs higher
computational capacities (see also Sect. \ref{sec:limitations}).

\subsection{Illumination pattern}
\label{sec:illpattern}
\begin{figure}[!ht]
\begin{center}
\includegraphics[width=7.cm, height=5.cm]{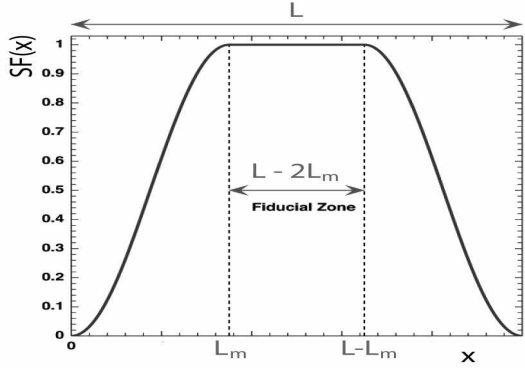}
\end{center}
\caption[] 
{\it
The smoothing function $SF(x)$. $L$ is the screen size, 
$L_m$ is the margin from the screen borders. 
}
\label{fidzone}
\end{figure}
To obtain the illumination pattern on the observer plane, we numerically compute the integral (\ref{hxy})
which can be written as a Fourier transform:  
\begin{eqnarray}
\label{intens}
h(x_{0},y_{0}) = \frac{1}{2\pi R_{F}^{\,2}} \Big\vert
FT\left[G(x_1,y_1)\right]\Big\vert ^{2}\,_{(f_x = \frac{x_0}{2\pi R_F^2},\,f_y = \frac{y_0}{2\pi R_F^2})} ,
\end{eqnarray}
where
\begin{eqnarray}
G(x_1,y_1) &=& exp\left[i\,\left(\phi(x_1,y_1)\,+\,\frac{x_1^{2}+y_1^{2}}{2R_{F}^{\,2}}\,\right)\right].
\label{Gxy}
\end{eqnarray}  
Here, ($x_1,y_1$) and ($x_0,y_0$) are the screen and observer coordinates, respectively. 
%Fourier transform is taken on 
%exponential of $\phi(x_1,y_1)$  (the generated phase screen) plus a quadratic term 
%$\,\frac{x_1^{2}+y_1^{2}}{2R_{F}^{\,2}}\,$.
Coordinates ($f_x,f_y$) are the conjugated variables in Fourier space. Before computing 
expression (\ref{intens}), a regularisation procedure for $G(x_1,y_1)$ has been defined to
avoid computational artefacts.

\subsubsection{Screen regularisation}
Since integral (\ref{intens}) is computed numerically, the coordinates ($x_1,y_1$) have discrete (integer) 
values describing pixel position centres on the screen, to
allow simple combinations of illumination patterns with different
pixel sizes (corresponding to different wavelengths).
%The program assigns the coordinates to the top left corner of the
%pixel. If we want to subtract the illumination pattern with different pixel size, we should centre the pixels coordinates.
%Moreover, as we used Fast Fourier Transform, the origin of ($f_x,f_y$) coordinates on observer plane is
%at top left corner of the image rather than the image centre. We thus should rearrange the frequency order
%of the illumination pattern.
That the integration domain is limited is physically
equivalent to computing the Fresnel integral within a diaphragm with the
size of the screen. In this case, we face a parasitic effect: the light
diffraction from the sharp edges of the diaphragm. This causes
rapid intensity variations at the borders of the observer plane. To
attenuate this effect and remove the resulting diffraction fringes, we
multiply the screen intensity transmission by a 2D smoothing function.
We define the following 1D smoothing function $S\!F(x)$ (see Fig. \ref{fidzone}):
%\[ S(x) = \left\{ \begin{array}{ll}
%\frac{1}{2}({sin}(\frac{3\pi}{2}-\frac{\pi x}{m})+1) & \mbox{$0\leq x \leq m$},\\[2mm]
%1 & \mbox{$m < x < L-m$},\\[2mm]     
%\frac{1}{2}({sin}(\frac{3\pi}{2}-\pi(1+\frac{x-L+m}{m}))+1) & \mbox{$L-m \leq x \leq L$},\\[2mm]
%0 & \mbox{{otherwise}},\\[2mm] 
%                           \end{array}
%                     \right. \]  
\[ S\!F(x) = \left\{ \begin{array}{ll}
\frac{1}{2}\left[1+\sin(\frac{\pi x}{L_m}-\frac{\pi}{2})\right] & \mbox{$0\leq x \leq L_m$},\\[2mm]
1 & \mbox{$L_m < x < L-L_m$},\\[2mm]     
\frac{1}{2}\left[1+\sin(\frac{\pi (x-L+L_m)}{L_m}+\frac{\pi}{2})\right] & \mbox{$L-L_m \leq x \leq L$},\\[2mm]
0 & \mbox{{otherwise}},\\[2mm] 
                           \end{array}
                     \right. \]  
where $L_m = 10 R_F$ is the margin length from the borders of the screen
with size $L$.
We multiply the function $G(x_1,y_1)$ by $S\!F(x_1) \times S\!F(y_1)$
in expression (\ref{intens}).
Since the Fresnel integral is dominated by the contribution of the integrand
within a disk that is within a few Fresnel radii, it is sufficient to smooth the discontinuity of
$G(x_1,y_1)$ within a distance of a few Fresnel radii (here $L_m = 10 R_F$).
%We controlled the effect of this procedure by checking the
%expected uniformity of the illumination from an uniform screen (\cite{TheseHabibi}).
%integral is due to 
%This choice of margin is made to 
We tested the efficiency of this regularisation procedure by
checking that the simulated illumination pattern from a point-source
projected through a uniform phase screen was --as expected--
also uniform beyond our precision requirements ($<1\%$) (\cite{TheseHabibi}).
%%(expected) flatness of a simulated illumination pattern
%%from an uniform phase screen
%%that with these margins we are able to reduce the intensity
%%perturbations due to border effects to a negligible level compared to
%%the scintillation fluctuations.
%%on the estimated intensity below .
%The final so-called fiducial zone is then a square
%with distance of 20$R_F$ from the borders of the observer plane.
To define a reliable fiducial domain {\it in the observer plane}
excluding the regions that are only partially illuminated owing to the diaphragm,
we also delimited an $R_{ref}/2$ margin from the borders of
the illumination pattern, corresponding to the typical radius (half-size) of
the large-scale luminous spots (see also \cite{colesetal}).
Figure \ref{pattern} shows the pattern produced by
a point source through a turbulent medium with $R_{diff} = 100\, km$
located  at $z_1=160\, pc$ from the Earth at wavelength $\lambda =
2.162\, \mu m$.
Since the corresponding Fresnel radius $R_F = 1300\, km$ is larger than the diffraction radius, 
the regime is the strong scintillation regime. The hot speckles (of
typical size $R_{diff}\sim 100\, km$) can be distinguished from the larger
dark/luminous structures that have a typical size of $R_{ref} \sim 2 \pi R_F^2/R_{diff} \simeq 
100,000\, km$.

\subsubsection{Effect of sampling}
Here, we discuss some limitations caused by the pixellisation. 
The screen should be sampled often enough to avoid the aliasing effects.
Aliasing happens when $G(x_1,y_1)$ contains frequencies that are higher than the Nyquist
frequency $f_{Nyq}$ = $1/(2\Delta_1)$, where $\Delta_1$ is the pixel size.
In relation (\ref{Gxy}), $G$ contains
two length scales, the diffraction and the Fresnel radii ($R_{diff}$
and $R_F$).
$R_{diff}$ is the characteristic length of the phase screen variations
$\phi(x_1,y_1)$; it is at least necessary that $\Delta_1 < R_{diff}/2$
in order to sample phase variations up to $1/R_{diff}$ spatial frequency.
%within the diffraction radius. 
$R_F$ appears in the quadratic term
$exp \left[i\,\frac{x_1^{2}+y_1^{2}}{2R_{F}^{\,2}}\right ]$; 
the oscillation of this
term accelerates as $x_1$ and $y_1$ increase, and aliasing occurs
as soon as the
distance between two consecutive peaks is smaller than 2$\Delta_1$. In one dimension
we therefore expect aliasing if
\begin{eqnarray}
\frac{(x_1+2\Delta_1)^2 - x_1^2}{2 R_F^2} > 2\pi, \nonumber \\ 
i.e.\ \frac{x_1}{\Delta_1} > \pi \left[\frac{R_F}{\Delta_1}\right]^2 - 1,
\label{alias1}
\end{eqnarray}
where $x_1 / \Delta_1$ is the distance to the optical axis, expressed in pixels.
%In terms of number of pixels $n_1$ = $x_1 / \Delta_1$:
%\begin{eqnarray}
%n_1 >  \pi (\frac{R_F}{\Delta_1})^2 - 1 . 
%\label{alias2}
%\end{eqnarray}  
The condition $\Delta_1 = R_F /2$ would be obviously insufficient
here to avoid aliasing, since beyond $\sim$ 11 pixels only from the optical axis 
%($\sqrt{x_1^2+y_1^2}$ > 11 pixel)
($x_1 / \Delta_1$ > 11 pixel),
the quadratic term would be under-sampled.
In practice, for
the configurations considered in this paper,
%infrared wavelengths and for a screen located at $\sim$ 100 pc,
the Fresnel radius is $> 1000\, km$.
%By chosing $\Delta_1 = 15\, km$ in the simulation,
%the aliasing starts at $x_1 / \Delta_1 > 13,900$ pixels from
%the centre of the image,
%corresponding to more than $200 R_F$
When $\Delta_1 = 22\, km$ in the simulation,
the aliasing starts at $x_1 / \Delta_1 > 6490$ pixels from
the centre of the image,
corresponding to more than $140 R_F$,
which is large enough to cover the sensitive domain related to
the stationary phase approximation.
%for a simulation
%with $N$ < 20,000, aliasing effect is usually avoided.
\subsubsection{Extended source (spatial coherency)}
The illumination pattern of a scintillating extended source is given
by the convolution product of the
illumination pattern of the point-like source
with the projected source limb profile (\cite{Moniez} \& \cite{TheseHabibi}),
\begin{eqnarray}
I_{ext} &=& \frac{L_s}{z_0^2} P_{r}\,*\,h,
\label{iprh}
\end{eqnarray}
where $L_s$ is the source luminosity, and the normalised
limb profile is described as a uniform disk:
%with sharp edges:
\[P_{r}(x_{0},y_{0}) = \left\{ \begin{array}{ll}
                          1/\pi R_s^2 & \mbox{$\sqrt{x^{2}_{0}+y^{2}_{0}} \leq R_s$}\\[2mm]
                          0 & \mbox{otherwise,}
                               \end{array}
                        \right. \]
where $R_s = \frac{z_{0}}{z_{1}}\,r_s$ is the projected source radius on
the observer's plane, and $r_s$ is the source radius.
Figure \ref{extended} shows the convolution of the pattern
of figure \ref{pattern} with the projected profile of a
star with radius $r_s$ = 0.5$R_{\odot}$, located at
$z_0$ ($160pc$) +$z_1$ ($1kpc$)=$1.16\, kpc$  ($R_s =53,000\, km$). 
%The geometrical configuration is  
%the same as figure \ref{pattern}.
High-frequency fluctuations due to diffractive speckle disappear, 
and the pattern loses contrast, but the variations on $R_{ref}$
scale remain visible.
%since $R_{ref} > R_s$.
As the convolution involves a disk of radius $R_s$, we can perform the calculation only 
at a distance greater than $R_s$ from the borders. We therefore define  a new fiducial zone by excluding a 
margin of $R_{ref}/2$ + $R_s$ from the initial borders. Any statistical analysis will be made within this zone
to be safe from any border perturbation.

\subsubsection{Polychromatic source (time coherency)}
The illumination patterns shown in Figures \ref{pattern} and \ref{extended} are computed for a 
monochromatic source 
(fixed $\lambda$), but observations are done through filters with
finite-width passbands. 
To take the contributions of different wavelengths to the pattern into account, 
we superimpose the illumination patterns obtained with the same refractive structure (the same 
column density fluctuations) at different wavelengths
\footnote{For a given physical screen characterised by the column density
$Nl(x_1,y_1)$, $R_{diff}$ varies with $\lambda^{6/5}$, as shown in
Eq. (\ref{relrdiffout}).}.
We have considered the passband of the SOFI camera in $K_s$ band 
and approximated it as a rectangular function over the
transmitted wavelengths with a central value 2.162 $\mu$m and  
width 0.275 $\mu$m.
Twenty-one illumination patterns were computed for 21 regularly spaced
wavelengths within the 
%different wavelengths were selected from the interval 
[2.08 , 2.28] $\mu$m interval, and were co-added to simulate the
illumination pattern through the $K_s$ passband.
We checked that the spacing between successive wavelengths was
small enough to produce a co-added image with a realistic
residual modulation index,
%--representative of the
%expectation from a uniform spectrum within the filter acceptance--
by studying this index as a function of the
number of monochromatic components equally spaced within the
full bandwidth. We found that
an asymptotic value is reached with a co-added image made up of only
about ten components.
%to compute their corresponding illumination patterns.

\begin{figure*}[!ht]
\begin{center}
\includegraphics[width=16.cm]{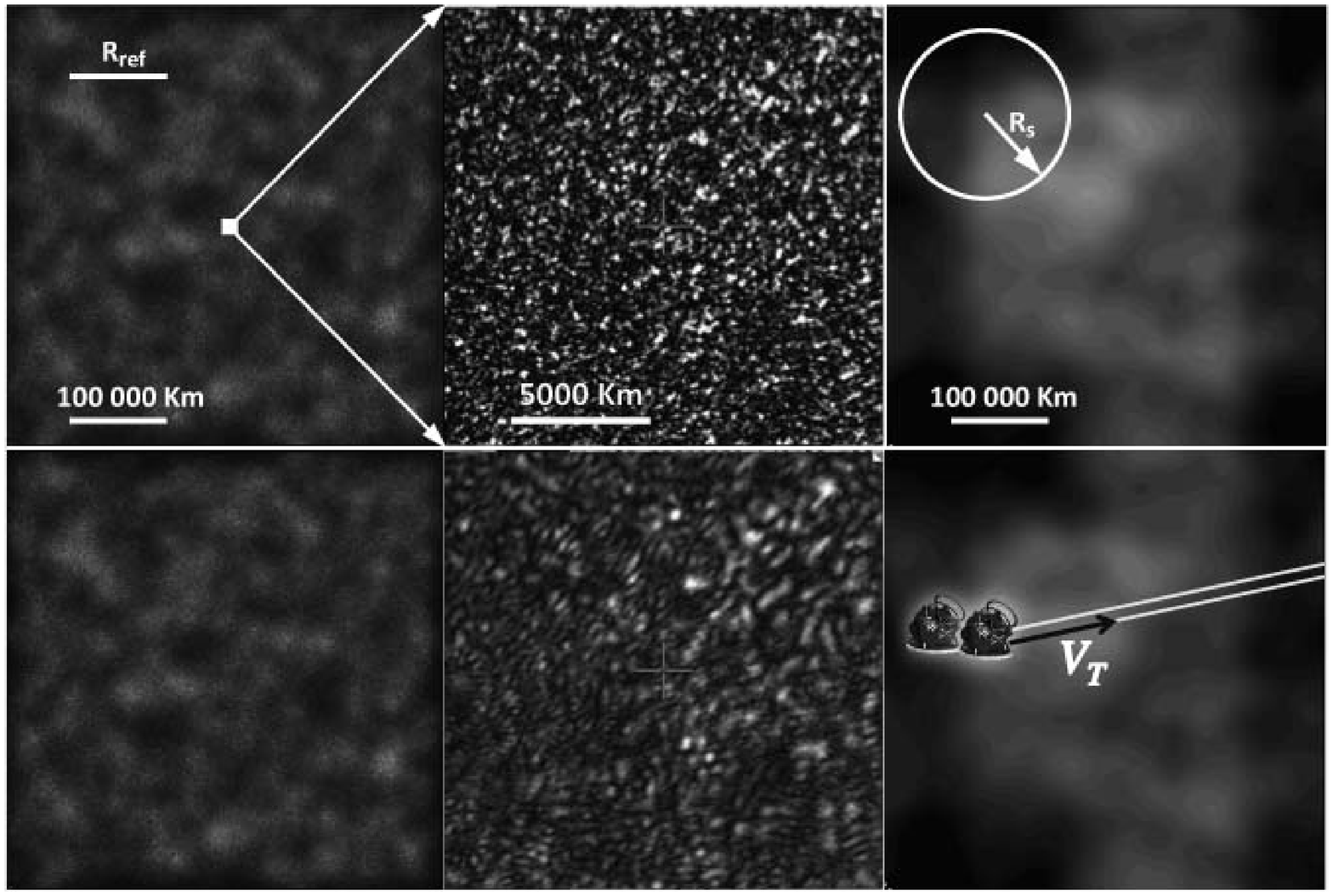}
\end{center}
\caption[] 
{\it
Simulated illumination maps ($20,000\times 20,000$ pixel of 
$22.6\,km$ side)
produced on Earth by a source located at $z_0+z_1=1.18\, kpc$
through a refracting cloud assumed
to be at $z_0=160\, pc$ with a turbulence parameter $R_{diff}(2.16\mu
m)=100\, km$. Here $R_{ref}(2.16\mu m) \simeq 100\, 000\, km$.

Top-left and middle: illumination produced at $\lambda=2.16\mu m$ from
a point-source with a zoomed detail; the contrast is $100\%$.
The grey scale ranges from $0$ to $4$ times the mean intensity.

Top-right: the same from a K0V star ($r_s=0.85R_{\odot}$, $M_V=5.9$,
at $1.18\, kpc$ $V=16.3$).
The circle shows the projection of the stellar disk ($R_S=r_s\times
z_0/z_1$). Here the modulation index is only $3.3\%$, and
the grey scale ranges from $\pm 20\%$ around the mean intensity.

The bottom maps are the illuminations in $K_s$ {\bf wide band}
($\lambda_{central}=2.162\mu m$, $\Delta\lambda = 0.275\mu m$),
using the same grey scales as above.
The modulation index is $55\%$ for the point-source (left and centre) and $3.3\%$ for
the extended source (right).

%Top left: Illumination pattern produced by a point-like monochromatic source. The contrast is 
%	100\%. Top right: The same from an extended source with 46\% contrast. Bottom left: The 
%      illumination pattern of a point-like polychromatic source. The contrast is 75\%.
%      Bottom right: The same for an extended polychromatic source shows 46\% contrast; the passband width does
%      not induce a significant deviation from the illumination pattern
%      of a monochromatic extended source.
%Each of the red lines materialise the relative trail of an observer
%within the illumination pattern. The observed light-curve is therefore
%given by the intersection of the line with the pattern.
The two parallel straight lines show the sections sampled
by two observers located about $10000\, km$ apart,
when the screen moves with the transverse velocity $V_T$.
}
\label{multypoint}
\end{figure*}

Figure \ref{multypoint} (left and centre) shows a comparison between
monochromatic (up) and 
$K_s$ passband (down) illumination patterns of a point-like source. 
The speckle pattern is attenuated when the light is not monochromatic
(or equivalently when time coherency is limited).
%and the modulation index decreases together with the time coherency.
%%The speckles are diluted and less contrasted.
%This can be understood by the fact that
%the size of the superimposed speckles varies by 
%\begin{equation}
%\frac{\Delta R_{diff}(\lambda)}{R_{diff}(\lambda)} = \frac{6}{5}\times
%\frac{\Delta \lambda}{\lambda} = \frac{6}{5} \times 10\% \sim$ 12$\%
%\end{equation}
%within the given wavelength interval, according to expression (\ref{relrdiffout}).
%This chromatic effect is due to
%the strong sensitivity of the constructive interference condition with
%the wavelength\footnote{
%Nevertheless, one should remember that the screen is a non dispersive medium
%for optical wavelengths (dielectric medium with an index independent of $\lambda$),
%contrary to the radioastronomy case (plasma), which
%results in a weaker chromaticity sensitivity.}.
%
This is caused by the small decorrelation bandwidth $\delta\lambda_{dec}$ of the strong
diffractive scintillation regime, which is given by
\begin{equation}
\delta\lambda_{dec} /\lambda \sim ( R_{diff}(\lambda)/R_F)^2
\end{equation}
(\cite{narayan92} 1992, \cite{gwinn}).
This chromatic effect results from
the high sensitivity of the constructive interference condition with
the wavelength\footnote{
Nevertheless, one should consider that the gaseous screen is a non-dispersive medium
for optical wavelengths (dielectric medium with an index independent of $\lambda$),
unlike the radioastronomy case (plasma), which may
result in weaker chromaticity sensitivity. This specificity deserves
more detailed studies, which are beyond the scope of this paper,
considering the minor impact of the bandwidth compared to the
smearing produced by the source size.}.
In the case of Fig. \ref{multypoint}, the decorrelation bandwidth is
$ \delta\lambda_{dec} /\lambda \sim 1\%$. Since the $K_s$
passband is $\delta\lambda /\lambda \sim 0.1 \sim 10\times
\delta\lambda_{dec} /\lambda$,
a first-order estimation of the modulation index for a $K_s$ passband pattern
is the value obtained when adding ten decorrelated patterns
of speckle {\it i.e.} one multiplied by $\sim \sqrt{1/10}$.
The modulation index found with our simulation ($\sim 55\%$) has
the correct order of magnitude.

By contrast, structures of size of $R_{ref}$ are much less sensitive to the variations
in $\lambda$,
according to the combination of (\ref{relrdiffout}) and (\ref{Rrefraction}) :
\begin{equation}
\frac{\Delta R_{ref}(\lambda)}{R_{ref}(\lambda)} = \frac{1}{5}\times
\frac{\Delta \lambda}{\lambda} = \frac{1}{5} \times 10\% \sim$ 2$\%.
\end{equation}
As a consequence of the large-scale smearing of the point-source
pattern when considering an \emph{extended} source, there is 
no significant difference between monochromatic and polychromatic
patterns for such an \emph{extended} source, as shown in Fig. \ref{multypoint} (right).
Therefore, in the following we ignore the impact of the $R_{diff}$ structures.
%The time coherence limitations affect the speckles but not the 
%large dark/luminous scales and for an extended source, the speckles disappear and only 
%the larger scales remain visible.
\subsection{Simulation of light curves}
\label{sec:simlightcurve}
What we observe with a single telescope is not the 2D illumination pattern but a 
light curve. Because of the relative motions, the telescope sweeps a 1D section of 
the pattern, at a constant velocity as long as parallax can be neglected.
We therefore simulated light versus time curves 
by sampling the 2D pattern pixels along straight lines, with the
relative speed of the telescope.
\subsection{Computing limitations}
\label{sec:limitations}
We adopted the same number of pixels $N\times N$
and the same pixel scale $\Delta_1$ to
numerically describe both the screen's and the observer's planes;
indeed, the light emerging from the screen
is essentially contained in its shadow, and this choice optimises the
screen filling.
Since the two planes are conjugated, the following relation arises:
\begin{equation}
N\Delta_1^2=2\pi R_F^2.
\end{equation}
Because of this relation, N has to be as large as possible to minimize
$\Delta_1$ and also to get a wide useful area. This area is indeed
restricted by the definition of the fiducial domain, which can be
heavily reduced when simulating the illumination from an extended
source. Also a large fiducial domain is essential when simulating
long-duration light curves.
As a consequence, memory limitations affect the maximum
size of the screen and illumination 2D patterns. In the present paper,
we were limited to patterns of $20,000\times 20,000$ pixels.

\section{Observables}
\label{sec:observables}
The observable parameters of the scintillation process are the
modulation index and the modulation's characteristic time scales.
The main observable we used in our data analysis (\cite{habibietal} 2011b) 
is the \emph{modulation index}. It is defined as  the flux dispersion
$\sigma_I$ divided by the mean flux $\bar I$: 
$m = \sigma_I/\bar I$.
We first compare the effective modulation index of simulated screens with the
theoretical expectations, then examine the precision we can reach
on $m$ from a light curve, and give a numerical example.
%We will describe the extraction of the
%light curves from the simulated illumination pattern and the way they represent the statistics of
%the original 2D pattern.
We also show how we can connect the observed modulation index to the
geometrical parameters ($R_{diff}(\lambda)$, $\lambda$, $R_s$ etc.) through simulation. 

\subsection{Modulation Index}
\begin{figure}[!ht]
\begin{center}
\includegraphics[width=8.cm]{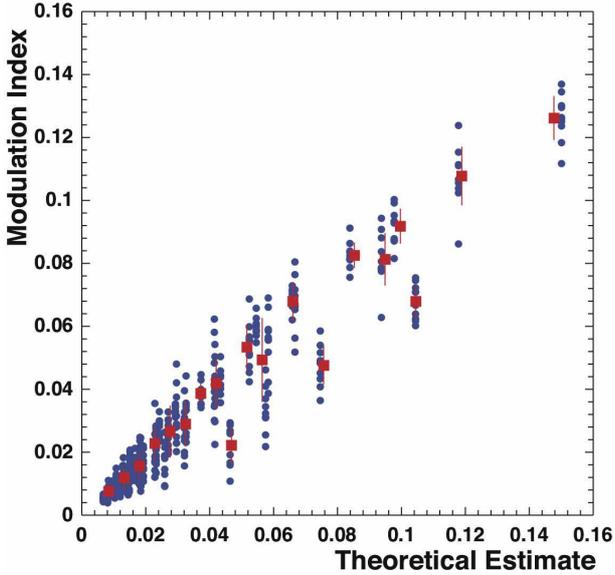}
\end{center}
\caption[] 
{\it
The effective intensity modulation index $m = \sigma_I/\bar I$ for simulated
%  scintillating stellar illumination patterns  as a function of $(R_F/R_s)^{1/3} (R_{ref}/R_s)^{5/6}$. 
scintillating stellar illumination patterns as a function of the
theoretical modulation index.
% $(R_{diff}/R_F)^{1/3} (R_{ref}/R_s)^{7/6}$. 
  The blue dots show the effective modulation index values for different realisations of
  the phase screens. The red squares represent the mean value of the effective modulation
  indices with the same
%$(R_{diff}/R_F)^{1/3} (R_{ref}/R_s)^{7/6}$
theoretical value. This plot shows the expected
  agreement for $m<0.15$, a domain where the simulated screens are large
  enough not to suffer from statistical biases (see text).
%allows us to constrain 
%  the screen parameters when knowing constraints on $m$.
}
\label{mx}
\end{figure}
For a {\it point-like} source in the strong scintillation regime the modulation index $m \approx 1$. 
For an {\it extended} source (radius $r_s$, projected radius
$R_s=r_s.z_0/z_1$) %, apparent radius $\theta_s=r_s/(z_0+z_1)$)
in the same regime, we always have $m < 1$, and \cite{narayan92} (1992) showed
that when the small-scale (diffractive) speckle is completely smeared,
%\[ m_e \approx \left\{ 
%\begin{array}{ll}
%(\frac{R_{diff}}{R_F})^{1/3} & \mbox{$R_s < R_{ref}$},\\[2mm]
%(\frac{R_{diff}}{R_F})^{1/3} (\frac{R_{ref}}{R_s})^{7/6} & \mbox{$R_s > R_{ref}$}.     
%\label{menarayan} 
%\end{array}
%\right. \]  
\begin{eqnarray}
%m_e &\approx& (2\pi)^{1/3}\, \left[\frac{R_F}{R_s}\right]^{1/3} \left[\frac{R_{ref}}{R_s}\right]^{5/6}.
m \approx \ \left[\frac{R_{diff}}{R_F}\right]^{1/3}
\left[\frac{\theta_{ref}}{2\pi \theta_s}\right]^{7/6}
\end{eqnarray}
in the case of the Kolmogorov turbulence.
Here, $\theta_{ref}=R_{ref}/z_0$ is the angular refraction radius
\footnote{Narayan uses $\theta_{scatt}$, which equals $\theta_{ref}/2\pi$. }
and $\theta_s$ the source angular radius. In this expression,
\cite{narayan92} (1992) assumed that $z_0 << z_1$, therefore
$z_0+z_1 \approx z_1$. We use the following
expression, which is formally identical to the previous one for
$z_0 << z_1$ but can also be used when $z_0$ is not negligible:
\begin{eqnarray}
m &\approx& \left[\frac{R_{diff}}{R_F}\right]^{1/3}
\left[\frac{R_{ref}}{2\pi R_s}\right]^{7/6} \ \\
 &\approx&  0.035 \left[\frac{\lambda}{1 \mu m}\right] \left[\frac{z_0}{1 kpc}\right]^{-\frac{1}{6}}
                      \left[\frac{R_{diff}}{1000 km}\right]^{-\frac{5}{6}} \left[\frac{r_s/z_1}{R_\odot/10 kpc}\right]^{-\frac{7}{6}}.
\label{xparam} 
%\label{menarayan} 
\end{eqnarray}

This relation can be qualitatively justified by noticing firstly that
$R_{diff}$ quantity has to be considered relatively to $R_F$ when considering
its impact on diffraction,
%characterises the variations of the first term in the
%integrand of \ref{} since
%$/R_F$ is the length where the stationnary phase approximation
%scale relatively to the natural Fresnel diffraction scale.
%The faster the first integrand of 
and secondly
that the convolution of the point-source pattern (see Fig. \ref{multypoint}-up)
with the projected stellar profile (following expression (\ref{iprh})) makes
the contrast decrease when the number of $R_{ref}$-size
domains within $R_S=r_S z_0/z_1$ increases.
%competes with the refraction radius $R_{ref}$ is the main.
The ``exotic'' exponents are related to the Kolmogorov turbulence.
It should be emphasised that this relation assumes the scintillation is
strong and quenched ($R_s/R_F > R_F/R_{diff} > 1$).

We checked this relation
%of $m_e$  with the $X$ factor, we performed
by performing series of
simulations with different phase screens and stellar radii. We generated series of screens with 
$R_{diff} = $ 50 km to 500 km by steps of 50 km. For each screen, we considered different sources 
--at the same geometrical distances-- with radii from 0.25$R_\odot$ to 1.5$R_\odot$ by steps of 
0.25$R_\odot$ and computed corresponding illumination pattern realisations.
The modulation indices were estimated within the fiducial zone for each 2D illumination pattern.
In figure \ref{mx}, the thus estimated $m$ for each generated pattern are
plotted as a function of the theoretical value expected
from expression (\ref{xparam}).
We note that the modulation has relatively large scatter. %for each $X$.
This is because the fiducial zone is
not large enough, and it contains a limited number of regions with sizes of
%$Max(R_{ref} ,R_s)$
$R_{ref}$
(the scale that dominates the light variations).
In some cases there are very few distinct dark/luminous regions (as in figure \ref{extended}). 
The number of such regions
within the fiducial domain is
%$N_R \sim \frac{d^2}{\pi \,  {max}(R_{ref},R_s)^2}$
$N_R \sim d^2/R_{ref}^2$,
where $d$ is the  
size of the fiducial zone. After discarding the cases with $N_R$ < 5,
we computed the mean and root mean square of the effective $m$ values,
%with $1\sigma$ error bars,
for each series of configurations
with the same theoretical modulation index.
Figure \ref{mx} shows that relation
(\ref{xparam}) is satisfied for a modulation index smaller than
$0.15$. % ($X<0.08$).
The higher values of $m$ are systematically underestimated
in our simulation owing to the limitation of the screen size,
which has a stronger impact on the number of large dark/luminous regions
in this part of the figure. This number is indeed smaller
(though it is larger than 5), and the chances for sampling the deepest
valleys
and the highest peaks are fewer, therefore biasing the modulation
index towards low values.
%After linearly fitting the simulated $m_e<0.15$ mean values
%(red squares in Fig. \ref{mx}) as a function of the 
%right hand side term of (\ref{menarayan}), and
%using relations (\ref{Rfresnel}) and (\ref{Rrefraction}),
%we find (XXXprobably to be changedXXX):
%
%
%\[ X = \left\{ 
%\begin{array}{ll}
%0.52 \, \left[\frac{\lambda}{1 \mu m}\right]^{-1/6} \left[\frac{z_0}{100 pc}\right]^{-1/6} 
%                      \left[\frac{R_{diff}}{100 km}\right]^{1/2} & \mbox{$R_s < R_{ref}$}, \\[2mm]
%0.34 \, \left[\frac{\lambda}{1 \mu m}\right] \left[\frac{z_0}{100 pc}\right]^{-1/6} 
%                      \left[\frac{R_{diff}}{100 km}\right]^{-5/6} \left[\frac{r_s/z_1}{R_\odot/10 kpc}\right]^{-7/6}
%                 & \mbox{$R_s > R_{ref}$}.     
%\label{xparam} 
%\end{array}
%\right. \]  
%\begin{eqnarray}
%m_e &=& 0.1 \, \left[\frac{\lambda}{1 \mu m}\right] \left[\frac{z_0}{10 pc}\right]^{-1/6} 
%                      \left[\frac{R_{diff}}{1000 km}\right]^{-5/6} \left[\frac{r_s/z_1}{R_\odot/10 kpc}\right]^{-7/6}.
%\label{xparam} 
%\end{eqnarray}

Therefore, by measuring the modulation index from an observed light curve
and using an estimate of star type ({\it i.e.} radius)
and distance $z_0+z_1\approx z_1$, we can constrain $R_{diff}(\lambda)$ from
this relation (\ref{xparam})\footnote{In (\cite{habibietal} 2011b), we used a simplified relation, with
  no significant impact on the resulting constraints.},
even with poor knowledge of the screen's distance $z_0$, considering
the slow dependence with this parameter.
This technique allowed us to infer constraints in \cite{habibietal} (2011b)
on the gas turbulence within galactic nebulae ($z_0 \sim 80-190\, pc$ and
$z_1=8kpc$), and upper limits on hidden turbulent gas within the
galactic halo (assuming $z_0\sim 10kpc$ and
$z_1+z_0=62kpc$).

% (\ref{xparam}) through figure \ref{mx}.

\subsection{Information from the light curves}
\begin{figure}[!ht]
\begin{center}
\includegraphics[width=8.cm]{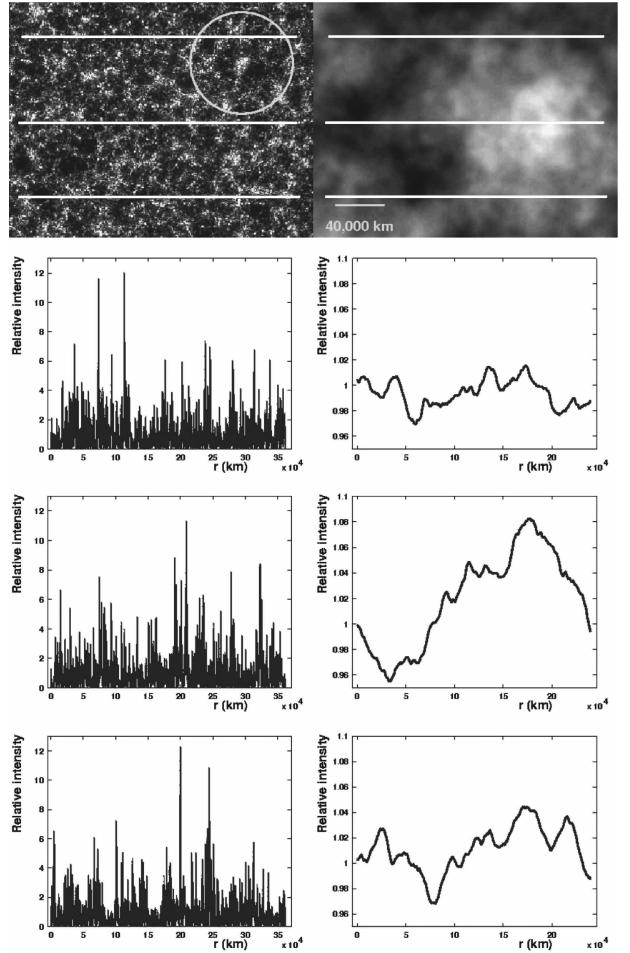}

\end{center}
\caption[] 
{\it
Light curves extracted along the 3 horizontal white lines for two illumination patterns. Left column: 2D pattern
	from a point-like source in $K_s$ band with $R_{diff}$ = 300 km %, $z_0$ = 125 pc, $z_1 \sim$
	%1 kpc 
	and $R_{ref} \approx$ 28000 km. The modulation indices of the three light curves differ
	by less than 5$\%$ from the 2D pattern modulation index. Right column: the illumination 
	pattern for an extended source with %$r_s$ = 0.5 R$_\odot$
        $R_s \approx$ 41000 km, through the same refractive screen. The 
	modulation indices fluctuate by more than $30\%$ around the 2D pattern index, implying 
	the necessity of longer light curves for a better statistical representativity. 
	The distance scale is common to both patterns. The circle
        shows the projected star disk.
%The luminosity scales (grey levels) are different for the two patterns.
The 3 light curves from the right column are not completely decorrelated,
because of their common proximity to the same large positive fluctuation.
}
\label{tlc}
\end{figure}
If a light curve is long enough (or
equivalently if the observation time is long enough), its series of light measurements 
represents an unbiased subsample of the 2D pattern. 
For instance, the left hand panels of figure \ref{tlc} represent the illumination pattern of a point-like 
source and three associated light curves. Here, $R_{ref}  \approx 28,000\, km$
and the modulation index $m_{point} = 1.18$. The light curves
are extracted from three horizontal parallel lines with lengths of $\sim 3.5 \times 10^5\, km$.
The corresponding time scale depends on the relative transverse velocity. 
The lines are selected far from each other in order not to be affected by the fluctuations of the same 
regions.
Modulation indices along the light curves differ from the 2D's by less
than $5\%$. When the modulation is
characterised by both $R_{ref}$ and $R_{diff}$, a light curve that spans a few $R_{ref}$
is a sample of the 2D screen, which is large enough to provide a good
approximation of the scintillation modulation index for a point-like source. 
The right hand panels of figure \ref{tlc} show the 2D pattern for an extended source with $R_s \approx 
41,000\, km$. Here, $R_s > R_{ref}$ and the flux fluctuations are smoothed, characterised by the 
unique length scale $R_{ref}$, and have a much smaller modulation index
$m_{extended} = 0.04$.
The light curves are extracted in the same way as the 
point-like source within the corresponding restricted fiducial zone (see figure \ref{extended}), 
therefore they span $\sim 2.5 \times 10^5\, km$ and statistically
include a little bit less than 10
$R_{ref}$-scale variations. Because of this statistically short length, the light-curve-to-light-curve
estimates of $m_{extended}$ typically fluctuate by $\sim 1/\sqrt{10} \approx 30\%$. The fluctuations on 
$m_{extended}$ estimates can only be reduced with longer light curves. 
%For a known cloud distance,
To get a precision value 
of $5\%$ on $m_{extended}$, the light curve should indeed be as long as $\sim 400\times R_{ref}$.
% $max(R_{ref},R_s)$.
As an example,
if we observe through a turbulent gaseous core with $R_{diff} = 200\, km$ in B68 nebula located at $z_0=80\, pc$
at $\lambda = 2.16\, \mu m$ ($R_{ref}\sim 27,000\, km$),
and assuming $V_T \sim$ 20 km/s, an
observing time of $\sim$ 150 hours %(XXXplease double check this numberXXX)
is needed to measure the modulation index with this precision of $5\%$.
When searching for unseen turbulent media 
located at unknown distances from us, the diffraction and refraction radii are unknown, and we can only
obtain a probability distribution of the observation time for a
requested precision on the modulation index.
\begin{figure*}[!ht]
\begin{minipage}[b]{0.55\linewidth}
\centering
%\vspace{3cm}
\includegraphics[width=10.cm]{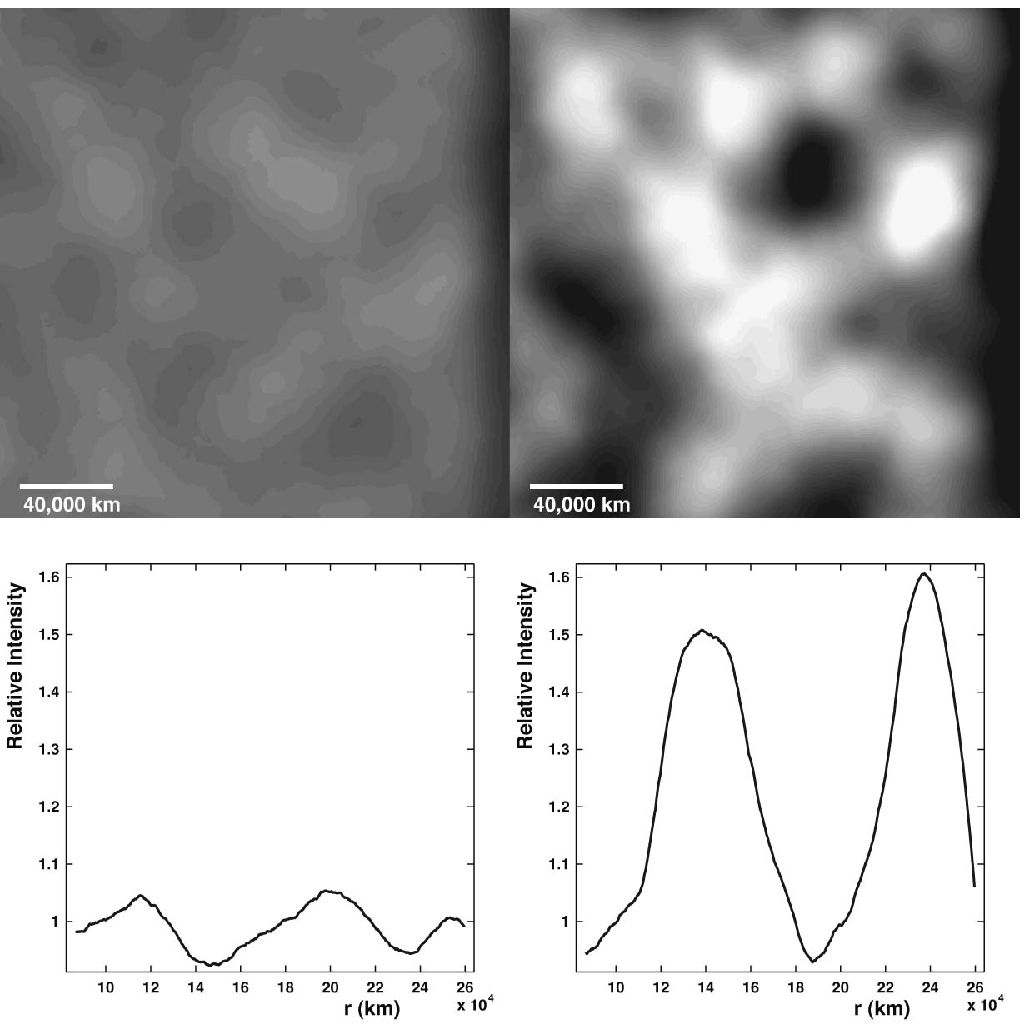}
\caption[] 
{\it
Simulation of illumination scintillation patterns with associated light curves      
from an extended source assuming a refractive screen with non-Kolmogorov
      turbulences (left $\beta=3.1$, right $\beta=3.9$, see text).
%Here the star radius 
%	is $r_s$ = 0.25$R_\odot$ ($R_S \sim$ 0.36 $R_{ref}$).
Here $R_{diff}$ = 100 km, $R_{ref} \approx$ 8 $\times$ 10$^4$ km
and the projected star radius is $R_S \sim 0.36 R_{ref}=28,800\, km$.
%The modulation index is more reduced for the pattern produced by
%	smaller turbulence index.
}
\label{ae}
\end{minipage}
\hspace{0.02\linewidth}
\begin{minipage}[b]{0.41\linewidth}
\centering
\includegraphics[width=7.cm]{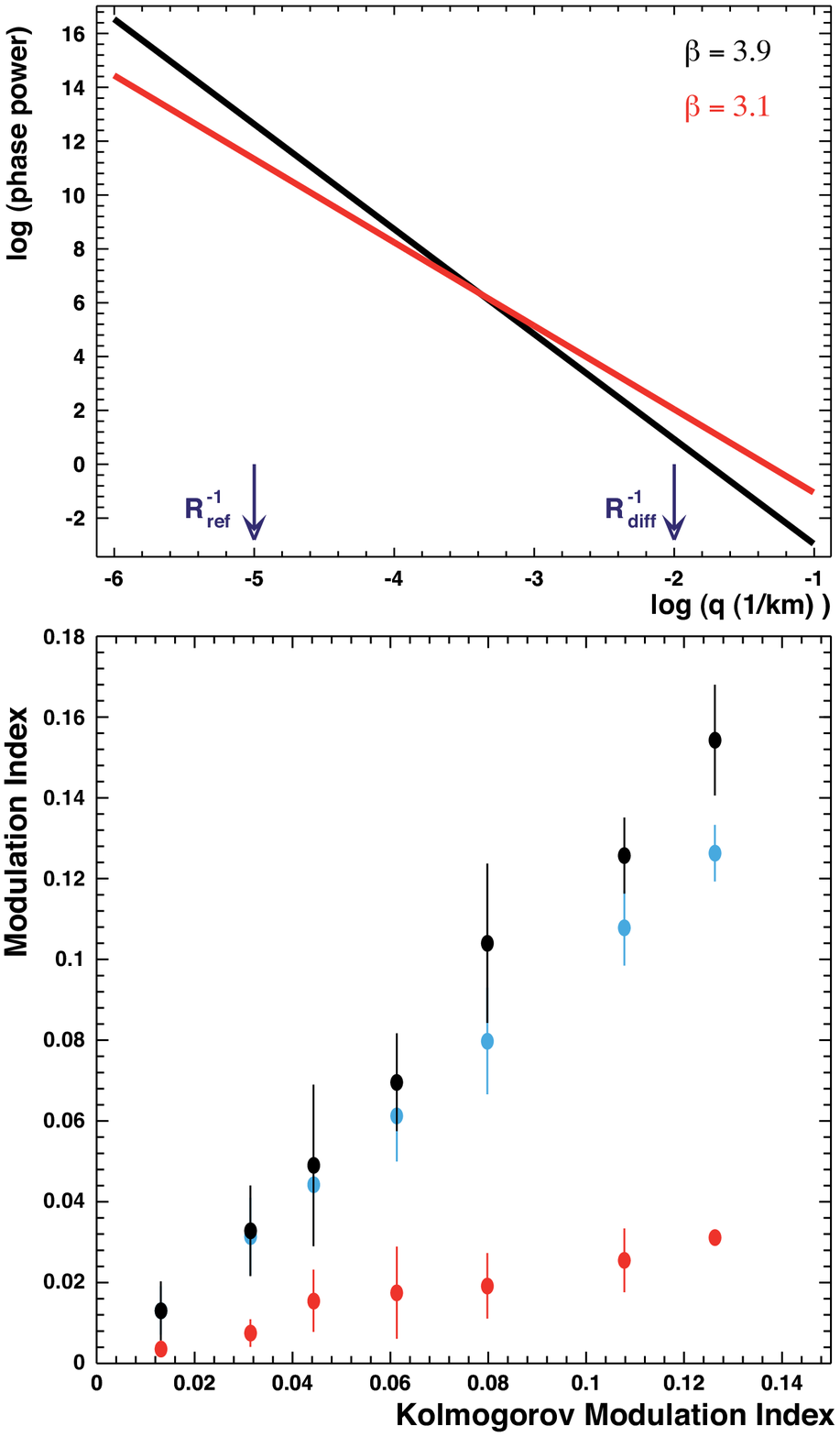}
\caption[] 
{\it
Top: Two different phase screen power laws.
Bottom: The corresponding modulation indices as a 
	function of the expected modulation index for the Kolmogorov
        turbulence.
The 3 indices plotted at a given abscissa correspond to screens with
$\beta=3.1$ (red), $3.67$ (Kolmogorov, blue, along the diagonal), and $3.9$ (black)
with the same $R_{diff}$.
% $X=(R_F/R_s)^{1/3} (R_{ref}/R_s)^{5/6}$
As the power law gets 
	steeper, a larger modulation is expected. Blue dots show
        the Kolmogorov turbulence case. 
}
\label{a3169}
\end{minipage}
\end{figure*}

The other important information carried by the light curves are the
characteristic time scale between peaks $t_{ref}=R_{ref}/V_T$, associated to the
refraction radius  (Fig. \ref{multypoint} upper left), a
correlation duration $t_{s}=R_{s}/V_T$, associated to the
source radius  (Fig. \ref{multypoint} upper right) and possibly
$t_{diff}=R_{diff}/V_T$, associated to the small diffractive speckle
structure (Fig. \ref{multypoint} lower left);
the latter could be detected
in exceptionally favourable cases (source with very small
angular size, assuming the
strong diffractive scintillation regime)
with a powerful detection setup such as LSST\footnote{
The detection condition would be $r_s/z_1 \lesssim 10\times
R_{diff}/z_0$
(for which the projected stellar disk includes less than $\sim 100$ speckle spots),
assuming a setup
able to sample the target star with $\lesssim 1\%$ photometric
precision every few seconds. The scintillation of an
$A0$ type star ($R=2.4 R_{\odot}$) in the LMC (magnitude $V=19.4$) seen
through a screen with $R_{diff}=100km$ at distance $30pc$ is an
example of such a favourable configuration, which could be discovered by
the LSST (\cite{LSST-science}).}.
The time power spectrum of the expected stochastic light curve should
show a slope break around frequency $t_{ref}^{-1}$, and possibly around
$t_{diff}^{-1}$ (\cite{goodman85}),
which should allow one to distinguish it from purely
randomly fluctuating light curves due to photometric noise, and
to extract constraints on the scintillation configuration.
Estimating $t_{s}$ would be challenging, since
the extended source acts as a low passband filter on the
point-source pattern; therefore, the imprint of the projected
radius is essentially
%not to be found in a peak of the time power spectrum
%(it is expected to attenuate existing peaks),
%in the typical
%interval between extremas,
%but in the smoothness of the light curve variations (or equivalently its
within the attenuation of the light curve autocorrelation
(\cite{goodman} 2006),
%the light-curve autocorrelation duration (that can be
%polluted by many experimental artifacts)
and more specifically --as mentioned above-- within the modulation index,
that can be polluted by many observational artefacts.
%In favorable cases (very small source), even the shorter time scale
%$t_{diff}=V_T/R_{diff}$ associated to the small diffractive speckle
%structure (Fig. \ref{multypoint} down-left) could be detected.
These potentialities of the time studies need more investigation and
should be discussed in more detail in a forthcoming paper.

\section{Probing various turbulence laws}
\label{sec:turbulence}
%We have now a good confidence in our simulation process.
Up to now, we
have focussed on the standard Kolmogorov turbulence.
In this section,
we investigate a possible  deviation from the Kolmogorov turbulence law\footnote{
For the observed supersonic turbulence (with $\beta$ > 11/3) see \cite{larson}.}.
We chose two different phase spectra with $\beta$ = 3.1 and
$\beta$ = 3.9 in relation (\ref{spectre}).
To study the corresponding scintillation modes, we generated two
series of phase screens according
to the spectra and computed the illumination patterns from an extended source 
%with $r_s$ = 0.25$R_\odot$ ($R_s \sim$ 0.36 $R_{ref}$) through each of them. 
with $R_s \sim$ 0.36 $R_{ref}$ through each of them. 
The patterns are represented in figure \ref{ae} with corresponding
light curve samples. 
For both images, $R_{diff}$ = 100 km, $R_F$ = 1150 km, and $R_{ref} \approx$ 8 $\times$ 10$^4$ km.
The pattern with $\beta$ = 3.1 (left) shows a small
modulation index $m(3.1)$ = 0.04, while the pattern with $\beta$ =
3.9 (right) shows a much larger modulation index 
$m(3.9)$ = 0.22.  As can be seen from its visual aspect, the
turbulence with a higher exponent produces stronger 
contrast on large scales compared to the other one ($\beta$ = 3.1).
To understand the origin of the difference, we compared the two phase
spectra at the top of figure \ref{a3169}.
The steeper spectrum ($\beta$ = 3.9) has more power for fluctuations on large scales. 
Moreover, by integrating equation (\ref{spectre}), we computed 
that the total power distributed from $R_{ref}$ to $R_{diff}$
is about an order of magnitude larger for $\beta$ = 3.9 than for $\beta$ = 3.1.
Therefore, the larger the $\beta$ exponent is the stronger flux fluctuations are produced. As 
a conclusion, the detection of scintillation should be easier for turbulences with steeper spectrum,
because a larger modulation index is expected.
%This is illustrated at the bottom of figure \ref{a3169} 
%where we plot the modulation indices produced by three screens with
%the same $R_{diff}$ (only changing with abscissa) for three
%different $\beta$ values (including the Kolmogorov case), as a function of the
%``classical'' Kolmogorov turbulence expected index.
This is illustrated at the bottom of figure \ref{a3169} 
where we plot the modulation indices produced by three screens with
different $\beta$ values (including the Kolmogorov case), as a function of the
``classical'' Kolmogorov turbulence expected index.
In this plot, each abscissa corresponds to a distinct $R_{diff}$
value, which is common to the three screens.
By increasing $\beta$,
we increase the modulation, especially for high $m$ values. 

\section{Discussion: guidelines provided by the simulation}
\label{sec:discussion}
The observations analysed in (\cite{habibietal} 2011b) were intrepreted
by using our simulation pipeline.
But we have used our simulation not only to establish connections between
the observed light curves and the scintillation configuration, but also
to define observing strategies as follows. \\
Firstly, a correct sensitivity to the scintillation needs the ability
to sample, with $<1\%$ photometric precision at a sub-minute rate, the
light curves of small distant background stars
($M\sim 20-21$), which have a projected radius small enough to allow
for a modulation index of a few percent (typically $< R_{\odot}\ at\ 10kpc$).
Our study of the time coherency shows that the usual large passband filters
can be used without significant loss of modulation index.
Since the optical depth of the process is unknown, a large field
of view seems necessary for the exploratory observations,
either toward extragalactic stellar sources within LMC or SMC
or through known gaseous nebulae.
To summarise, an ideal setup for searching for scintillation with
series of sub-minute exposures
would be a $\sim 4\, m$ class telescope equipped with a fast readout
wide field camera and a standard filter (optical passband to
search for invisible gas towards extragalactic sources, infrared to
observe stars through visible dusty nebulae). \\
Secondly, our work on the simulation provides us with a guideline to find
an undisputable signature of scintillation.
The first possibility consists in the search for chromatic effects.
Subtle chromatic effects
betwen the different regimes (associated to different time scales)
have been shown, but will probably be hard to observe.
Figure \ref{multypoint} (left and centre) shows the expected speckle image from
a point source. The position/size of the small speckles
(characterised by $R_{diff}(\lambda)$) are very sensitive
to the wavelength, and it is clear that a desynchronisation of the
maxima would be expected when observing such an idealised point source
through different (narrow) passbands. But that real sources
have a much larger projected radius than the speckle size
completely screens this chromatic effect. The impact of the
wavelength on the position/size
of the wide (refractive) spots (Fig. \ref{multypoint}-right)
is much weaker (following $\lambda ^{-1/5}$ according to the
combination of expressions (\ref{relrdiffout}) and
(\ref{Rrefraction})). Therefore, only a weak chromatic effect
is expected from an extended source, even when observing with
two very different passbands.
For a clear signature of scintillation,
it seems easier to take advantage of the rapidly varying
luminosity with the observer's position within the illumination pattern.
As a consequence, simultaneous observations
with two $\sim 4\, m$ class telescopes
%(needed to get sub$\%$
%photometric precision in sub-minute exposures
%for small distant stars of $M\sim 20-21$)
at a large separation (few $10^3\, km$)
would sample different regions of the illumination patterns
(see Fig. \ref{multypoint} bottom right), and
therefore measure (at least partially) decorrelated light curves as shown in
Fig. \ref{tlc} (right).
The decorrelation will be complete if the distance between the
two telescopes is greater than $2\times max(R_{ref}, R_s)$.
A single observation of
such a decorrelation will be sufficient to definitely confirm the discovery of
a {\it propagation} effect that cannot be mimicked by an intrinsic variability.

\section{Conclusions and perspectives}
Through this work, we have simulated the phase delay induced by
a turbulent refractive medium on the propagation of a wave front.
We discussed the computational limitations of 
sampling the phase spectrum and to obtaining large enough
illumination patterns.
These limitations will be overriden in the near future
with increasing computing capabilities.
The illumination pattern on the observer's plane has been computed for the
promising strong regime of scintillation, and the
effects of the source spatial and time coherencies have been included.    
We have established  the connection between the modulation
index (as an observable) of the illumination pattern
with the geometrical parameters of the source and
the strength of the turbulence (quantified by $R_{diff}$).
Furthermore, we showed that when the spectral index of
the turbulence increases (as in the case of supersonic turbulence),
the detection of scintillating light curves should be easier. \\
%Determining the refraction radius $R_{ref}$ is another way to estimate the 
%diffraction radius of the turbulent cloud.
These simulation studies and, more specifically, the modulation index
topic were successfully used in our companion paper
(\cite{habibietal} 2011b)
to interpret our light curve test observations of stars located behind
known galactic nebulae and of stars from the Small Magellanic Cloud,
in the search for hypothetical cold molecular halo clouds. \\
Time scales, such as $t_{ref}$ = $R_{ref}/V_T$ and $t_s$ = $R_s/V_T$ (where $V_T$ is the relative
velocity between the cloud and the line of sight), are observables 
that we plan to study further and in detail. Their extraction can be done
through analysing the time power spectrum 
of the light curves, and it should give valuable information on the
geometrical configuration, as well as on the turbulent medium.
%Although in weak
%regime of scintillation and in strong regime with $t_s$ > $t_{ref}$,
%the expected modulation
%index is small but the projected stellar radius $R_s$ remains a characteristic length scale which may leave
%a signature on the time spectra of the light curves.\\

The observing strategy has been refined through the use of the
simulation, and we showed that observing desynchronised
light curves simultaneously measured by two distant four-metre class telescopes
would provide an unambiguous signature of
scintillation as a propagation effect.

\begin{acknowledgements}
We thank J-F. Lestrade, J-F. Glicenstein, F. Cavalier, and P. Hello
for their participation in preliminary discussions. We wish to thank
our referee, Prof. B. J. Rickett, for his very careful review, which helped
us to significantly improve the manuscript.
\end{acknowledgements}

\end{document}